
\magnification=1200
\tolerance 10000
\parindent 0pt
\parskip 0.2cm
\baselineskip 0.5cm
{\voffset 0.1cm
{\centerline{\bf{Microlensing events of the LMC are better explained}}}
{\centerline{\bf{by stars within the LMC than by MACHOs}}}
\vskip 0.5cm
{\centerline{ Kailash C. Sahu }}
{\centerline{Instituto de Astrofisica de Canarias, }}
{\centerline{38200 La Laguna, Tenerife, Spain.}}
{\centerline{  E-mail: ksahu@ll.iac.es}}
\vskip 0.6cm
{\centerline{{\bf To appear in}}}
\vskip 0.1cm
{\centerline{{\bf Pub. Astron. Soc. Pacific (Sept. 1994 issue)}}
}
\bigskip
{\bf{\centerline{ABSTRACT}}}
\smallskip
The recently reported microlensing events of the LMC have caused much
excitement, and have been interpreted as due to `dark objects' (MACHOs)
in the halo of our Galaxy. It is shown here that the stars within the
LMC play a dominant role as gravitational lenses and can indeed account
for the observed events.

For observations within the bar of the LMC, the probability
of microlensing being caused by a star within the LMC is
found to be $\sim$  5 $\times$ 10$^{-8}$.
Outside of the bar, the probability of  microlensing being
caused by a star in the LMC is 4 to 12 times lower. The MACHO event
(Alcock et al. 1993) and one of the EROS events (Aubourg et al. 1993)
lie within the bar for which the probability of microlensing is
consistent with being caused by an object within the LMC.

If the stars within the LMC play a dominant role as lenses,
the events should be concentrated towards the center of the LMC.
On the other hand, if MACHOs play a dominant role as  lenses then,
for a given number of monitored stars, the events should be
uniformly distributed over the whole area of the LMC.
Thus the galactic and the LMC lenses can be statistically
distinguished in most cases.

It is further shown that, under certain conditions, the light curve
of an event caused by a star within the LMC would be
different  from the one caused by a MACHO.
This can also be a distinguishing signature, and seems to have been
observed in case of the MACHO event.
The fit discrepancy near the peak which the
authors say ``is not yet understood" would be a natural consequence if the
event is caused by an object within the LMC, which further indicates that
the lensing is due to a low mass star within the LMC itself.
\medskip
{\centerline{***********}}
\vfill \eject
{\centerline{\bf{1. Introduction}}}

Most galaxies are known to have flat rotation curves (Begeman, 1987;
Sanders and Begeman 1994).
In the outer parts of the galaxies in particular, the
visible matter falls far short of what is required to explain the flat
rotation curves.
In our own Galaxy, the rotation curve is observed to be flat up to
at least 16 kpc from the center (Fich et al. 1989).
The scenarios proposed to explain the flat
rotation curves are either departure from Newtonian dynamics
in large scale, or presence of dark matter (for a review see
Sanders 1990 and Broeils 1991).
In the latter and more conventional
scenario, the rotation curves would imply that a significant part of the
  mass of the galaxies, including our own,
resides in the halo in some form of dark matter.
The nature of this dark matter has been hypothesized to be in either of
the two forms: MACHOs (i.e. massive compact halo objects which is a
collective
term for `Jupiters', brown dwarfs, red dwarfs, white dwarfs, neutron stars
or black holes); or elementary particles such as massive neutrinos, axions or
WIMPs (weakly interacting massive particles such as photinos etc.).
One important distinguishing factor
of the MACHOs is that they can have detectable gravitational effects
and can cause ``gravitational microlensing" of background stars.

Einstein (1936) was the first to point out that a star can act as a
gravitational lens for another background star, if the two are sufficiently
close to each other in the line of sight. Given the observational
capabilities of that time, Einstein had however considered this to be a
purely theoretical exercise and had remarked that there was ``no hope of
observing such a phenomenon directly".

The observational capabilities have improved remarkably since then.
More recently,  Paczy\'nski (1986) worked out the probability of such
microlensing events by MACHOs and showed that if the halo of
our Galaxy is made up of
MACHOs, the probability of finding them through microlensing
is 5 $\times$  10$^{-7}$, independent
of their mass distribution. He suggested an experiment to look for such
events using the LMC stars. Further details of this idea
were worked out by Griest (1991). Such an experiment was taken up by 2 groups
who have reported their first results (Alcock et al. 1993;  Aubourg et al.
1993).
\smallskip

{\centerline{\bf{2. The experimental results}}}

The results from the MACHO experiment (Alcock et al. 1993) are
from the monitoring  of 1.8 million stars
in the bar of the LMC, both in blue and red wavelengths, for a period of
one year, which led to the detection of one event. More recently, it has been
reported that the analysis of all the data, which involves
monitoring of 8.3 million stars in the region of the bar over 2 years,
has led to the detection of 4 events (Cook et al. 1994).
The EROS experiment (Aubourg et al. 1993) involves
2 programs: one involves monitoring of 8 million stars over a field of
5 $\times$ 5 degrees using Schmidt plates in red and blue,
with a sampling rate of one or two per night
over a period of 3 years. The second program involves monitoring of 100.000
stars in red and blue wavelengths, over an area of 1 $\times$ 0.4 square
degrees with a sampling rate of about half an hour,
for about 7 months. The analysis of 40\% of data of the first program led to
the detection of 2 events, and
no event was detected in the second program.
The details of the published events are listed in Table 1.
 \smallskip
{\centerline{\bf{3. Concerns if the microlensing is caused by MACHOs}}}

If these are genuine microlensing events by galactic halo objects,
this could potentially herald an end to the problem of dark matter.
But this has also been the cause for some concern:  if these events are
caused by the objects in the halo, it would lead to some problems in
terms of stellar evolution and the theory of galaxy formation (Hogan, 1993).

Recent work on deuterium abundance derived from the observations
of a quasar suggests that the dark halos of galaxies may well be
nonbaryonic (Songaila et al. 1994).

In case of the MACHO event,  which has the highest signal to noise ratio,
there is a discrepancy between the best fit and the observed
light curve near the peak of the light curve. This consistently repeats
both in the blue and in the red,  and according
to the authors this ``fit discrepancy near the peak is not yet understood".
 Furthermore, the rate of
microlensing events as observed seems to be lower than expected from the
dark halo, and it is important to estimate the rate expected from the stars
that are known to exist in our galaxy and in the LMC.  The analyses for the
galactic stars and the LMC halo have been done, but the importance
of the LMC stars had been overlooked so far.
Sahu (1994) has recently shown that the stars within the LMC indeed play a
dominant role as gravitational lenses, the full details of the
probability analysis is presented here. It should be
noted here that, for the microlensing events of the galactic bulge stars
(Udalski et al. 1993; Udalski et al. 1994), the stars within the bulge
also play an important role as gravitational lenses, the analysis of
which has recently been done by Kiraga and Paczy\'nski (1994). Furthermore,
this paper presents the analysis of the light curve and shows that
if a star within the LMC acts as a lens, the source can be extended
which reproduces the light curve better.
Before attempting to analyse the light curve in detail,
let us first see in detail the probability of this scenario.

\smallskip
{\centerline{\bf{4. Analysis of the probability}}}

{\bf 4.1 Mass of the bar and stellar density}

To calculate the probability of the microlensing being caused by a star in
the LMC itself, we need to know the stellar mass density in the LMC.
Let us first confine ourselves to the bar of the LMC. From the surface
luminosity, it is estimated that the observed luminosity
of the bar is about 10 to 12 \% of the total observed luminosity of the LMC
in the optical wavelengths (De Vaucouleurs and Freeman 1973,
Bothum and Thompson 1988).
To see whether the effect of extinction is important and to
account for it, we note that the extinction in the
region of the bar, as derived from the 100 $\mu$m IRAS maps
is between 12 and 24 MJy per steradian with an average
value of about 17 MJy per steradian (Schwering and Israel 1990), which
translates to an average extinction of about 1.5 mag in the V band
(Laureijs, 1989). This value is
consistent with the value deduced by Hodge (1991) from background galaxy
counts and is also consistent with the value derived from the observations of
the most reddened stars in the region (Isserstedt 1975). The
average extinction in the outer regions is  about 0.3 to 0.4 magnitudes.
It is perhaps worth noting here that, as we will see later, errors in
extinction measurements have only very little effect on the final probability.
To see how the extinction affects the mass, let us assume that
the extinction is uniform in depth, and let $d$ be the total depth.
Thus for any line of sight, if A$_v$ is the total extinction, the
ratio of the observed to the true luminosity at that point
$${L_{obs}\over{L_{true}}} = {1\over d} \int_0^d{2.5^{-A_v z/d} dz}
 =  {1 - e^{-0.916 A_v}\over{0.916 A_v}}\eqno (1)$$
where $z$ is the direction of the line of sight.

Using  A$_v$=1.5 mag for the bar, and 0.3 to 0.44 mag for the region outside,
we get
$$ \left[ {L_{obs} \over L_{true}}\right] _{bar} = 0.54 \ and \
 \left[ {L_{obs}\over L_{true}}\right] _{out} = 0.82 \ to  \ 0.87$$
Now, using the fact that
${\left( L_{obs}\right) _{ bar}} / {\left( L_{obs} \right)_{ tot}}$
= 0.1 to 0.12, and after a little algebra we get,
$${\left( L_{true}\right) _{ bar} \over \left( L_{true}\right) _{ Tot}} =
0.14 \ to  \ 0.18$$
Thus the true luminosity of
the bar is about 14 to 18 \% of the total luminosity of the LMC. The total
mass of the LMC, as determined from
various methods, ranges from 6 x 10$^9$ to 15 x 10$^9$ M$_\odot$
(De Vaucouleurs and Freeman, 1973). (In case of the LMC,
there is no significant discrepancy between the mass derived from the
observed luminosity and the mass derived from the velocity dispersion.)
Assuming that the mass to light ratio in
the bar and the outer parts are the same, we get the mass of the bar to be in
the range 1 x 10$^9$ M$\odot$ to 2.7 x 10$^9$ M$\odot$. In our subsequent
calculations, we will round off the mean and use  2 x 10$^9$ M$\odot$ as the
mass of the bar. Considering the fact that the LMC
is gas poor  and only about 5\% of the LMC
mass is neutral hydrogen (Westerlund 1990, Israel and de Graauw 1991,
Rohlfs et al. 1984),  we will neglect the contribution
of gas to the mass and assume this entire mass to be made up of stars.

{\bf 4.2 Expected probability of microlensing in the region of the bar}

To calculate the probability of microlensing, let us assume that there are
N$_{tot}$ stars being monitored. We note that, the stars that are farther
away, have more stars in front of them, thus having a larger probability of
being microlensed. But the observed number density of stars becomes less
at larger depths because of extinction. Thus, in order to correctly calculate
the probability, we must know how the extinction affects the observed number
of stars with depth. To make life easier, we can take the current monitoring
programs and the magnitude limits to calculate the probability. We note that
the limiting magnitude of the current surveys are about 20 to 21, which at
the
distance of LMC, corresponds to an absolute magnitude of 1.5 to 2.5. So  the
magnitude of the stars that can be observed at the near side is 1.5 to 2.5
whereas the magnitude of the stars at the far end is 1.5 magnitude brighter.
If the distribution of stars among different spectral types is assumed to
be similar to
what is observed in the solar neighborhood, then the difference in stellar
number density from front end to the back is about 3 (Allen 1973).
Similar results are found from HST observations of LMC clusters (Gilmozzi
et al. 1994). (If the
extinction is 3 magnitudes, then this value is about 10. As it turns out,
the effect of extinction is small. If the extinction is larger, the
mass of the bar becomes larger thus increasing the probability of
microlensing. But the extinction effectively reduces the observed number of
stars which are at a larger depth which have the larger probability of being
microlensed, thus decreasing the probability. These 2 effects compensate for
each other the net effect is that the probability decreases with extinction,
but only slightly. If the extinction is 3 magnitudes instead of 1.5,
the net probability changes only by less than 50\%. Furthermore,
at the magnitudes considered here, the age is not expected to play a
significant role, so to a good approximation, the spectral distribution
of the stars can be assumed to be similar to that in the solar neighborhood
for the probability calculation.)

Assuming the extinction to be uniform in depth, we can express the observed
number of stars at any layer $dz$, at a depth of $z$, as

$$N_{obs}(z) = {N_z} 3^{-{z\over d}} dz \eqno (2)$$
where $N_z$ is the observed number of stars per unit depth in absence of
extinction i.e, the observed number of stars per unit depth at the near side.
If $N_{tot}$ is the total number of stars being monitored, then
$$N_{tot} = \int_0^d{N_z 3^{-{z\over d}} dz} = N_z d \left[ {1-1/ln 3\over
ln3}\right] = 0.6 N_z d \eqno (3)$$
Substituting eq. 3 in eq. 2, we get
$$N_{obs}(z) \simeq  {N_{tot}\over 0.6d}  {3^{-{z\over d}} dz}\eqno (4)$$
The fraction of area covered by the Einstein rings af all  the individual
stars lying in the front of this layer can be expressed as
$$A_f(z) = \int_0^z{\pi R_E^2(z) n(z)  dz} \eqno (5) $$
where $n(z)$ is the stellar number density at depth $z$. Since $n = \rho /m$
and $R_E^2 \propto m$ where $\rho$ is the stellar mass density and $m$ is
the mass of the star
$$ A_f(z) = {4\pi G \over c^2}\int_0^z{\rho (z) z dz} \eqno (6)$$
which is thus independent of the mass distribution. Assuming the stellar
density to be uniform with depth

$$A_f(z) =  {2 \pi G \rho z^2 \over{c^2}}\eqno (7) $$

Fom Eq. 4 and 7, the instantaneous probability of observing one microlensing
event, when $N_{tot}$ stars are being observed, is
$$ P = \int_0^d{N_{obs}(z) A_f(z) dz} =  {2 N_{tot} \pi G \rho \over 0.6 d c^2}
\int_0^d{3^{-{z\over d}} {z^2
}}dz\eqno (8)$$

Strictly speaking, this is valid for $P<<1$, which is the case here. The
optical depth $\tau$, which is equivalent to the probability of any star
being microlensed at a given time is $P/N_{tot}$,  is easily
obtained by solving eq. 8 as

$$\tau \simeq {0.5 \pi G \rho  d^2 \over c^2} \eqno (9)$$

For the value of $\rho$, we can substitute,

$$\rho = {M  \over {L W d}} \eqno (10)$$
where L, W, d and M are the length, width, depth and mass of the bar
respectively.

Assuming L=3000pc, W=d=600pc and M = 2 $\times$ 10$^9$ M$\odot$,
and substituting eq.10  in eq.9,  we get,
$$\tau_{bar} \simeq 5 \times 10^{-8}\eqno (11)$$

with some uncertainty due to the uncertainties in the size and mass of
the bar. It must be mentioned here that the value of the optical depth
calculated above is an average value for the whole bar. Since the
luminosity gradient in the bar itself is about 30\% from the center
to the outer parts (Bothum and Thompson 1988), the optical depth
would be slightly higher in the central region of the bar, and
slightly lower in the outer region.

It is clear from Eqs. 9 and 10 that $\tau \propto d$. In absence of any more
accurate information, we have used the depth to be the same as the width and
have neglected any inclination effect. In the literature one finds that the
bar can be as thin as the disk itself (Binney and Tremaine 1987). So,
considering the fact that the thickness of the LMC disk is about 300pc,
and the inclination is about 45 degrees (Westerlund 1991), the depth can
be about 450pc in the extreme case, thus the effect due to the uncertainty
in $d$ can be  25\% at the most.

Carrying out an identical analysis for the region outside the bar
it is easy to see that, if the depth of the LMC disk is
between 100 to 300pc (Feast, 1989), the optical depth in the region outside
the bar is about 4 to 12 times smaller.

{\bf{4.3. Comparison with observed probabilities}}

The MACHO event  lies in the central region of the bar.
Keeping in mind the uncertainties involved in
the estimation of optical depth on the basis of the single published
MACHO event, we can only say that it
seems to be well below the optical depth of
$5 \times 10^{-7}$  expected from a dark halo made up entirely of
MACHOs, and is consistent with  the optical depth calculated above.
Extending an identical analysis for the more recent results of 4 events
reported from the monitoring of 8.3 million stars over a period of 2 years,
the estimated observed probability is found to be indeed very close to the
value
calculated above, although there is some uncertainty in the value
of the detection efficiency.
In case of the EROS events, one event lies in the outer region of the bar,  the
other is far from the bar, and the optical depth has been estimated to be
higher ($\sim 2 \times 10^{-7}$, Gould et al. 1994).
But Gould et al. (1994) also suggest that there may be some systematic effects
and the detection efficiencies may be uncertain. The situation will be
clearer as more events are observed, particularly events with higher
magnifications (see below), and we must wait for more events to be
observed before making any direct comparison with the calculated
optical depth.

Given the fact that the number involved in the statistics is small,
and the fact that the calculations are rather straight forward and do not
involve any unknown quantities or quantities with large uncertainties, the
observed and the expected probabilities are certainly not inconsistent with
each other.

{\bf{4.4. Optical depth to microlensing by objects other than the LMC stars}}

The probability of microlensing due to stars in the disk and halo of our own
galaxy has been discussed by Gould et al. (1994) and the possible contribution
of LMC halo was discussed by Gould (1992).
The probability of
the events being caused by the expected number of white dwarfs in the halo
also appears to be too small. Apart from other theoretical reasons, the
mass of the objects derived from the observed events do not agree well with
this being caused by the the white dwarfs in the halo since they
 have a mass distribution which is highly peaked around
0.6 M$_\odot$ (Weideman 1990).
 \smallskip
{\centerline{\bf{5. Can we observationally distinguish between a MACHO and a
star in the LMC?}}}

This probability analysis may be enough for most astronomers to abandon
the idea of MACHOs since the microlensing can be effectively caused by
the known stellar population in the LMC itself and one does not need to
resort to unknown objects such as MACHOs for an explanation.
However, let us go one step further and investigate whether
there is a distinguishing feature which can
tell us whether the microlensing is caused
by a MACHO or a star in the LMC.
We will now see that under certain
conditions, the fit discrepancy near the peak of the light curve
can indeed be a distinguishing feature of the microlensing
being caused by an object within the LMC.
Let us see this in detail.

The amplification sharply rises  when the
lensing object comes close to being perfectly aligned with the source.
The physical reason for this is that when the lensing object and
the source are perfectly aligned, the image becomes a ring (also
called the Einstein ring) instead of
two separate images and thus the amplification rises sharply.
 [For details, see Paczy\'nski, 1986]. In the case of lensing by MACHOs,
$R_E$ is of the order of 1.2 $\times$ 10$^{14}$ $\sqrt{M\over M_\odot}$ cm,
assuming $D_d$ = 10 kpc and $D_s$ = 55 kpc which, at the source plane, is
$\sim$10$^{15}\sqrt{M\over M_\odot}$ cm. The diameter of the source is thus
2 to 3 orders of
magnitude smaller even if the lensing object is $\sim$0.1M$_\odot$. Thus the
probability of the source and the lens being perfectly aligned is extremely
small, and can happen only for events with extremely small $u$ ($^<
\hskip -0.3truecm _\sim \hskip 0.3truecm 0.01$),
which is unlikely and certainly not the case in the observed events.
However, the value of $R_E$ projected to the source plane can be much smaller
if the lensing object is close to the source. For example, if we consider a
typical distance of about 100 pc between the source and the lensd, R$_E
\simeq  10^{13} \sqrt{M\over M_\odot}$ cm, which, for a 0.1
M$_\odot$ lens, is only about about 3 times smaller than the typical
radius of a red giant. Thus, in order that at least a part of the source
is perfectly aligned with the lensing object, the impact parameter has to be
$^< \hskip -0.3truecm _\sim \hskip 0.3truecm $0.3, which may  explain the
fit discrepancy near the peak of the light curve.

Note also that for the maximum amplification to be appreciable, the source
size must not be too large for a different reason. If it is too large,
even if both objects are perfectly aligned, most of the source is still not
perfectly aligned with the source, this reduces the maximum amplification.
This is  given by (Paczy\'nski 1986)
$$ A_{max} = [1 + {4R_E^2\over{r_{0}^2}}]^{1\over{2}} \eqno (12)$$

{\bf{5.1 Light curve due to an extended source}}

Assuming the source to be extended, let us calculate the amplifications.
A general methodology for such a case was developed by Bontz (1979).
The geometry of our particular case however (shown in Fig. 1)
enables us to develop a simpler solution, which is found to be
more suitable for numerical integrations.
A mathematical approximation and the resulting analytical solution
for such a case has also been given
 by Schneider et al. (1992)
which, unfortunately for our purpose, does not have sufficient accuracy
just at the transition point where the fit discrepancy begins to occur.
Nevertheless, this analytical expression
was found to be very useful in providing a constant check on the correctness of
the
numerical algorithm used here to calculate the light curve.

In a general case, the amplification caused by an extended source would be
given by (Eq. 6.81 of Schneider et al. 1992)
$${\int d^2y \ I(y) \ \mu_p(y) \over \int d^2y \ I(y)} \eqno (13)$$
\noindent where I(y) is the surface brightness profile of the source,
$\mu_p(y)$ is the amplification of a point source at point
$y$, and the integration is carried out over the entire surface
of the source.

Let us assume the source to be a disk of uniform brightness with
radius $r_0$  and let  $u$ be the impact
parameter (=$l/R_E$). $y_0$ be the distance between the lens and the
center of the source projected onto the lens plane. Let us choose a circular
coordinate system and let ($r,\theta$) be the representative
point which is at a distance $y$ from the lens.
{}From Fig. 1 we see that

$$y = \sqrt{y_0^2 + r^2 - 2 \ y_0 \ r \ cos(\gamma )} \eqno (14)$$
where $$y_0=\sqrt{x_0^2 + l^2},\ \gamma = \pi -\theta -\alpha \ and \  \alpha
= sin^{-1}(l/y_0) \eqno (15)$$
i.e.$$y = \sqrt{ y_0^2 +r^2 +2 \ y_0 \ r \ cos(\alpha+\theta )} \eqno(16)$$
The amplification can be written as
$$\mu_p(y) = {y^2 + 2 \over y(y^2+4)^{1/2}}  \eqno (17)$$
Thus from eq.(13) the amplification can be expressed as
$$ {{\int_0^{r_0} \int_0^{2\pi} {y^2+2\over y(y^2+4)^{1/2}} \
y  \  dr \ d\theta} \over { \int_0^{r_0} \int_0^{2\pi}  y \ dr
\ d\theta}} \eqno (18)$$
This is an expression suitable for numerical integration and can  be
easily integrated. The light curve can be reproduced as a function of time
$t$, which is related to $x_0$ through the relation $x_0 = (R_E/t_0)t$ where
$t_0$ is the timescale of microlensing
(which is the time taken by the lens to cross its own Einstein ring).
$y$ in turn, is related to $x_0$ through eq.15 and 16.

Since I myself do not have the original data,
 I digitised the published
light curve, in order to facilitate my attempts on fitting the light curve.
Hence the error bars are not shown here, although the error bars at the
points near the peak were also carefully digitised.
The light curve was  numerically calculated
for different values of $u$ and $r_0$
to reproduce the observed  curve.
The best fit to the light curve is obtained for $u=0.17$ and
the value  of r$_0/R_E$ between 0.17 and 0.175,
which is shown in Fig.2.
The solid curve is the best fit obtained using the lens to be a MACHO,
and the dotted curve is obtained assuming the lens to be a star within the
LMC.
As is clear, the curve due to a  MACHO has a significant
discrepancy at the peak of the light curve both in the blue and in
the red which is reduced in case of the best-fit
light curve  with a star in the LMC as the lens.

For a $\chi^2$ analysis, let us first consider the red
light curve, the observations of which have a higher signal  to
noise ratio. The total $\chi^2$ gain, defined as
$\Sigma\{ [A(obs)-A(MACHO)]/\sigma\} ^2
- \{ [A(obs)-A(LMC star)]/\sigma\} ^2$ where A refers to the amlification,
is about 7.3 units if we consider the 4 points near the peak. For the blue
light curve, which has a lower S/N, the total $\chi^2$ gain for these 4
points is about 4 units. For other points, both curves are practically
identical.

It must be emphasized here that the source is not always extended for an
LMC lens. The source would be point like for smaller values of $r_0/R_E$
(which would be the case if the  mass of the lens
is slightly higher, or, the distance between the
source and the lens is slightly higher). As shown in the next section, the
solution becomes fairly unique if the
light curve can be reproduced by an extended source approximation.

\smallskip
{\centerline{\bf{6. Mass of the lenses and further confirmation of the
scenario}}}

If this extended source approximation is valid, we can use the value of
$u$ and r$_0/R_E$ as calculated in the preceeding section and proceed to
calculate some physical parameters.
The source, from its color and luminosity, has been inferred to
be a clump giant, for which we can assume the radius to be $r_0=7 \times
10^{11}$
cm. Now,
$$ {r_0\over R_E}  = {r_0\ c \over \sqrt{4GMD}} \simeq 0.17; D \simeq D_{ds}
\eqno (19)$$
i.e.$$\sqrt{{M\over M_\odot}{D\over 100pc}} \simeq 0.3 \eqno (20)$$
The mass the lens can also be expressed as
$$M \simeq {[t V_e c]^2 \over {16GD(1-u^2)}} \eqno (21)$$

{}From Eq. 20 and 21, which now provide a consistency check, we get
$$ \sqrt{{M\over M_\odot}{D\over 100pc}} = 0.44 \left({V_e\over 40 km s^{-
1}}\right) \simeq 0.3 \eqno (22)$$
Thus, irrespective of the value of M and D for this particular case, we get the
value of $V_e$ to be $\sim 27$ km s$^{-1}$. The fact that this is similar to
the value of dispersion obtained for the bar from radio observations
(Rohlfs et al. 1984) further confirms the validity of this scenario.
The mass of the lens can thus be expressed as
$$M \simeq 0.1 {100pc\over{D}} M_\odot \eqno (22)$$

Although the signal to noise ratio in the EROS events might
not allow us to do such a detailed analysis of the
light curve, the observed amplifications would imply
that the source can be approximated as a point source.
For the point source approximation to be valid, we need
$r_0/R_E < 1$ (Fig. 1) which provides a lower limit to the mass of the lens.
An upper limit to the mass of the lens comes from equation 21 (using the
velocity dispersion). The mass range thus derived is shown in Table 1.

\smallskip
{\centerline{\bf{7. Conclusions}}}

No dark matter
is required to explain the observed microlensing events of the LMC stars.
The probability of microlensing  being
caused by a star in the bar of the LMC is found to be consistent with the
observations.
Detailed analysis of the light curve of the event with the best signal to
noise shows that the light curve is better reproduced  by a
star in the LMC than by a MACHO, and indicates that a low mass star
in the LMC must be the cause of the microlensing.

The LMC induced events should be strongly clustered towards the central
region of the LMC. (They should be proportional to the number of monitored
stars multiplied by the integrated stellar mass density along the line of
sight.) In  case of the the galactic events, for
a given number of monitored stars, the events  should be uniformly distributed
over the whole of LMC. (They should be
simply proportional to the number of monitored stars.) Furthermore, under
certain conditions, the observed light curve in
case  of the LMC event can be  different from the light curve due to a
galactic lens. Thus the two scenarios can be distinguished statistically in
most cases, and individually in a few cases.
Such a distinguishing feature seems to have already  been
seen in case of the light curve of the MACHO event.

Harvey Butcher, Paul Hodge and Bengt
Westerlund promptly replied to my queries which were  helpful in clarifying
some  doubts regarding the structure
of the LMC. The author is grateful to Bohdan Paczy\'nski for his very
encouraging and useful remarks.
The help of Enrique Perez in the digitisation of the light curve was
valuable. Some details of the observational program
provided by Ken Freeman are gratefully acknowledged.

\smallskip
{\parskip 0pt
{\centerline{\bf{References}}}
\smallskip

Allen, C.W., 1973, {\it Astrophysical Quantities}, 3rd Edition,
The Athlone Press

Alcock, C. et al. 1993, Nature, {\bf 365}, 621

Aubourg et al. 1993, Nature, {\bf 365}, 623

Begeman, K., 1987, Ph.D. thesis, Groningen University

Binney J., Tremaine, S.,  1987, Galactic Dynamics, Princeton
University Press

Bontz, R.J., 1979, Astrophys. J., {\bf 233}, 402

Bothum, G.D., Thompson, I.B., 1988, Astron. J., {\bf 96}, 877

Broeils, A., 1991, Ph.D. thesis, Groningen University

Cook, K., et al., 1994, Bull. Am. Astron. Soc., {\bf 26}, 912

De Vaucouleurs, G, Freeman, K.C., 1973, Vistas Astron. {\bf 14}, 163

Einstein, A., 1936, Science, {\bf 84}, 506

Feast, M., 1989, ``Recent Developments in Magellanic Cloud
Research", Observatoire de Paris (eds
de Boer, K.S., Spite, F. \& Stasinska, G.) 75

Fich, M, Blitz, L, Stark, A., 1989, ApJ, {\bf 342}, 272

Gilmozzi, R., Kinney, E.K., Ewald, S.P., Panagia, N., Romaniello, M., 1994,
Preprint

Gould, A. 1992, ApJ, 404, 451

Gould, A., Miralda-Escud\'e, J., Bahcall, J.N.,
1994, ApJ Lett., {\bf 423}, L105

Griest, K, 1991, ApJ, {\bf 366}, 412

Hodge, P., 1991, IAU Symp. 148 ``The Magellanic Clouds", Ed.
R. Haynes and D. Milne, Kluwer Academic Publishers, p57

Hogan C.J., 1993, Nature, 365, 602

Israel, F.P., de Graauw, Th., 1991, IAU Symp. 148 ``The Magellanic Clouds", Ed.
R. Haynes and D. Milne, Kluwer Academic Publishers, p45 \par

Kiraga, M., and Paczy\'nski, B. 1994, ApJ (in press)

Kroupa, P, Tout, C.A., Gilmore, G., 1993, MNRAS, {\bf 262}, 545

Isserstedt, J.,  1975, Astron. Astrophys. {\bf 41}, 175

Laureijs R., 1989, Ph.D. thesis, Groningen University, p85

Paczy\'nski, B, 1986, ApJ, {\bf 304}, 1

Rohlfs, K., Kreitschmann, J., Siegman, B.C., Feitzinger, J.V., 1984,
Astron. Astrophys. {\bf 137}, 343

Sahu, K.C., 1994, Nature, {\bf 370}, 265

Sanders, R.H., 1990, Astron. Astrophys. Rev., {\bf 2}, 1

Sanders, R.H., and Begeman, K. 1994, MNRAS, 266, 360

Schneider, P, Ehlers, J. and Falco, E.E., 1992, ``Gravitational Lensing",
published by Springer-Verlag.

Schwering P.B.W., Israel, F.P., 1990,  ``Atlas and Catalog of Infrared
Sources in Magellanic Clouds", Kluwer Academic Publishers.

Songaila L.L., Cowie, C.J., Hogan, C.J., Rugers, M.,
1994, Nature {\bf 368}, 599

Udalski, A., et al. 1993, Acta Astr., 43, 289

Udalski, A., et al. 1993, Acta Astr., 44, 165

Weidemann,V., 1990, Ann. Rev. Astron Astrophys., {\bf 28}, 103

Westerlund, B., 1990, Astron. Astrphys. Rev., {\bf 2}, 29

Westerlund, B.,  1991, IAU Symp. 148 ``The Magellanic Clouds", Ed.
R. Haynes and D. Milne, Kluwer Academic Publishers, p15 }

\vfill \eject
{\bf{\centerline{Figure captions}}}

{\bf Fig. 1} Schematic of the geometry of the microlensing
used for calculation of the light curve.
The lens is assumed to be a point lens.
Note that the Einstein ring radius associated with an LMC
lens is much smaller than either a disk or a halo lens.
Consequently, the minimum impact parameter $u$ ($=l/R_E$)  needed for
a part of the source to be perfectly aligned with the
lens can be higher. This implies that the source can be resolved
at much smaller magnifications if the lens resides in the LMC.
See text for the details of the symbols used.

{\bf Fig. 2.} The observed microlensing events both in blue and red.
The best fit due to a MACHO is shown by a solid curve. The best fit
due to a star within the LMC as the lens is shown as the
dotted curve. (See text for details.)
\vskip 0.5cm
{\centerline{\bf TABLE 1}}
{\centerline{Details of the microlensed stars and the derived masses of the
lenses}}
\medskip
\settabs 7 \columns
\hrule
\medskip
\+Event  &  Magnitude & Amplification& Amplification & Event
 & Impact& Mass &\cr
\+& (m$_v$) & (in blue)&(in red)&
duration (d)\quad \quad \quad  & parameter&  (M$_\odot$)&\cr
\medskip
\hrule
\medskip
\+MACHO&  19.6 & $\sim$7.7 &   $\sim$7.7& 33.9&
0.17& 0.1 &\cr
\medskip
\+EROS 1&   19.3 &  1$\pm$0.1mag& 1$\pm$0.1mag&
44& 0.425& 0.002--0.32 &\cr
\medskip
\+EROS 2&    19.3  &  1.1$\pm$0.2mag&
1.3$\pm$0.2mag & 53& $\sim$0.34&0.0002--0.44& \cr
\medskip
\hrule

\vskip 1cm
\noindent The masses are derived under the assumption that the lensing
is caused by stars within the LMC. Note that the lower limit to the mass of
the lens comes from the requirement of a point source as explained in
section 6. This is only a lower limit
and consequently, the actual mass can be much higher than this limit.
On the contrary, the upper limit comes from the fact that
the tangential velocity of the lens cannot be much higher than
the velocity dispersion (Eq 21). This is an exact expression
and considering the usual distribution of the velocity
dispersion, the mass is more likely to be closer to the upper limit.
For the first event
of Aubourg et al. we have used the radius  appropriate
for a subgiant and for the second event the radius appropriate for a
normal star as indicated by their colours and luminosities.
For a  distance D between
the source and the lens, the mass has to be multiplied by 100pc/D.
For the MACHO event, the amplifications and the impact parameter
are from the extended source approximation.

\bye